\documentclass{article}

\usepackage{arxiv}

\usepackage[utf8]{inputenc} 
\usepackage[T1]{fontenc}    
\usepackage{hyperref}       
\usepackage{url}            
\usepackage{booktabs}       
\usepackage{nicefrac}       
\usepackage{microtype}      
\usepackage{lipsum}
\usepackage{graphicx}
\usepackage{color,array,amsthm}
\usepackage{grffile}
\usepackage{cite}  
\usepackage{amsmath,amssymb,amsfonts}
\usepackage{textcomp}
\usepackage{wrapfig}
\usepackage{verbatim}
\usepackage{mathtools}
\usepackage[utf8]{inputenc}
\usepackage{algorithm2e}
\usepackage{algpseudocode}
\usepackage{subcaption}
\usepackage{caption}

\title{Threat-Based Resource Allocation Strategy for Target Tracking in a Cognitive Radar Network}

\author{
 Ji~Ye~Lee \\
 Graduate School of Military Digital Convergence \\
 Ajou University\\
 Suwon 16499, South Korea  \texttt{jkllffdsa@ajou.ac.kr} \\
   \And
 J.H Park \\
 Graduate School of Military Digital Convergence \\
 Ajou University\\
 Suwon 16499, South Korea
 \texttt{parkjo05@ajou.ac.kr} \\
}

\begin{document}
\maketitle
\begin{abstract}
Cognitive radar is developed to utilize the feedback of its operating environment obtained from a beam to make resource allocation decisions by solving optimization problems. Previous works focused on target tracking accuracy by designing an evaluation metric for an optimization problem. However, in a real combat situation, not only the tracking performance of the target but also its operational perspective should be considered. In this study, the usage of threats in the allocation of radar resource is proposed for a cognitive radar framework. Resource allocation regarding radar dwell time is considered to reflect the operational importance of target effects. The dwell time allocation problem is solved using a Second-Order Cone Program (SOCP). Numerical simulations are performed to verify the effectiveness of the proposed framework. 
\end{abstract}


\section{Introduction}
Radar network is a system in which multiple radars work together and communicate with each other using information collected from the target~\cite{b24}\cite{b26}\cite{b27}. Currently, based on the development of communication technology, radar networks have gained growing attention. Since tracking data can be utilized in a coordinated way, there are some benefits of radar networks compared to a conventional single-radar operation, such as an increased coverage area and increased accuracy of target tracking and detection~\cite{b1}\cite{b25}. For example, the integrated information from the radar network can be used to detect Low Probability of Interception (LPI) target in a military system. 

In radar networks, research on resource management is an essential issue since the information shared through channels is affected by the allocated resources of each node. That is, the accuracy of the obtained information may vary depending on how the limited resources are used, even with the same geographical arrangement of the radar nodes~\cite{b4}\cite{b15}\cite{b16}\cite{b17}\cite{b18}. 
Liu~et~al.~\cite{b2} aimed to increase target state estimation accuracy and lower the false alarm rate of tracking a target in a cluttered environment. To solve this problem, Collaborative Detection and Power Allocation (CDPA) scheme was developed, and the optimization problem was solved by using the Cramer--Rao Lower Bound (CRLB) as the analytical cost. 
On the other hand, Yan~et~al.~\cite{b3} derived an optimization problem for multiradar--multitarget situations by considering which target to allocate to each radar and how to allocate the limited time resources of each radar to the target.

In some fields of radar networks, a threat-based approach is proposed to calculate threats using target state parameters and use this information for scheduling. Such a threat-based resource management technique has the advantage of an operational concept that can be efficiently applied to a scheduling algorithm~\cite{b5}\cite{b6}\cite{b11}\cite{b13}\cite{b14}\cite{b23}. For the case of air defense phased-array radar, Zhang~et~al.~\cite{b7} designed an appropriate threat model of the target so that the prior information of the target could be fully utilized. Additionally, by designing a 2D dynamic priority table, the threat level of the target and the deadline of the task were considered comprehensively to calculate the priority. Furthermore, Li~et~al.~\cite{b12} introduced a target prioritization algorithm based on three-way decision theory. Pang and Shan~\cite{b8}\cite{b10} calculated the target missing risk and the sensor radiation interception risk in the target tracking and detecting scenario. Once the risk was specified, resource management was conducted with the goal of minimizing the defined risk. Unlike the aforementioned works focused on reducing risk, Katsilieris~et~al.~\cite{b9} proposed a method to increase the threat accuracy of the target, where the operator was defined according to the operational purpose.

To learn the environment and plan the next action in a human-like way, cognitive control systems have recently been actively studied~\cite{b28}~\cite{b30}~\cite{b31}~\cite{b32}~\cite{b29}. In~\cite{b19}, a perception--action cycle, which is a distinguishing feature of a cognitive dynamic system, was mathematically modeled. 
This cognitive control system was used in a radar system to build cognitive radar.
In~\cite{b21}, Predicted Conditional Cramer--Rao Lower Bound (PC-CRLB), an appropriate evaluation matrix for predicting the tracking performance in a cognitive radar environment, was developed, and the optimization problem was solved using a SOCP.
However, these recent cognitive radar studies only considered the target state among the full set of features of the environment. In reality, not only the target state but the operational risk, such as the safety of assets, should be considered, especially in the case of air defense radar. In a battlefield situation, the accurate operational risk of the target is a key requirement for operators.

From this point of view, we specifically develop two factors in this study.
We propose a general condition that can accommodate the operator's requirements, and we mathematically verify an example case of the condition. In other words, a cognitive radar resource management model that includes the threat of the target is presented. The developed model is much closer to actual use requirements, and operational flexibility is also increased.
In addition, a new gain function is proposed to better express the risk-based performance. Subsequently, the performance is verified using numerical simulation. 

This paper is organized as follows.
In Section II, the necessity of a threat-based approach in radar systems is discussed. Then, a risk-based approach framework for cognitive radar is presented in Section III. In Section IV, a dwell-time allocation example is given, and it is validated via numerical simulation in Section V. Finally, the conclusion is given in Section VI.

\section{The necessity of a threat-based approach in radar systems}

According to~\cite{b21}, an object that threatens a valuable asset can be defined as a threat, and the loss due to this threat can be defined as a risk. In a military environment, a flying enemy object can be considered a potential threat. Threats can be labeled with numbers or letters in consideration of kinetic information as well as other features such as target type and intent. This series of processes is called threat assessment. Contrary to risk, which is a predicted loss after an event occurs, this labeled value represents the current relationship between the observed target and the asset.
In this study, the threat has the same meaning as the threat label~\cite{b9}, and it is defined as a value between 0 and 1.

\begin{figure}
\centerline{\includegraphics[width=\textwidth]{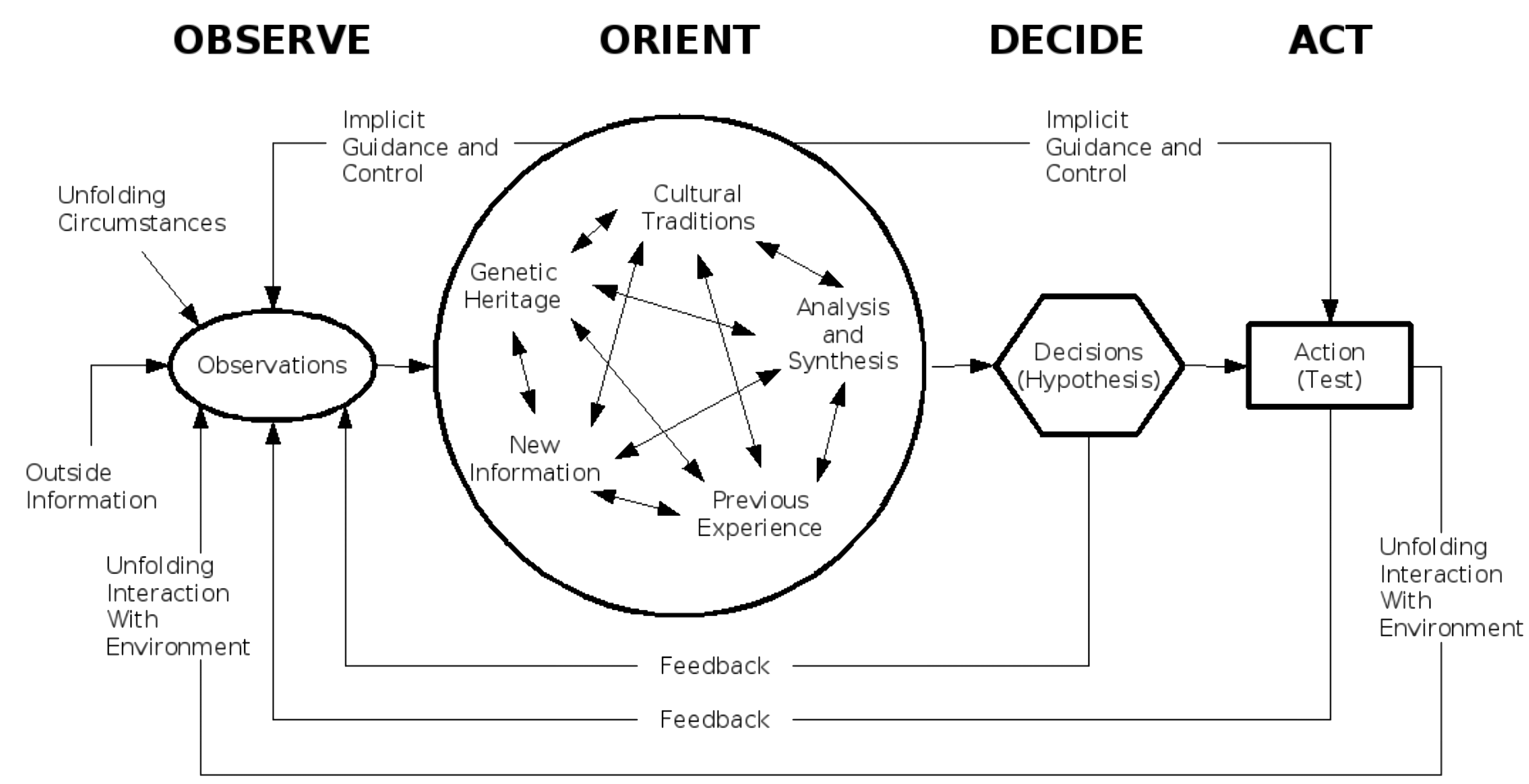}}
\caption{Observe--Orient--Decide--Act (OODA) loop cycle used to make decisions~\cite{b23}.}
\label{ooda} 
\end{figure}

The Observe-Orient–Decide–Act (OODA) loop shown in Fig.~\ref{ooda} is a well known decision making process in the military domain, and this method can be applied in an antimissile environment~\cite{b23}. An observed raw measurement should be assessed so it can be used to make tactical decisions, and threat assessment can be seen as an example of such a process. In the real world, threat assessment is performed by a human observing a target. However, if the threat can be determined by processing only kinetic information, the probability distribution of the threat can be mathematically calculated using the probability distribution of the kinetic information~\cite{b9}. Then, the operator can decide whether to attack the threat by considering the obtained threat label value and the resources of the currently available weapon. Since the assessed threat label value affects the determination of engagement, the accuracy of the obtained threat label has operational importance.

\begin{figure}
\centerline{\includegraphics[width=30pc]{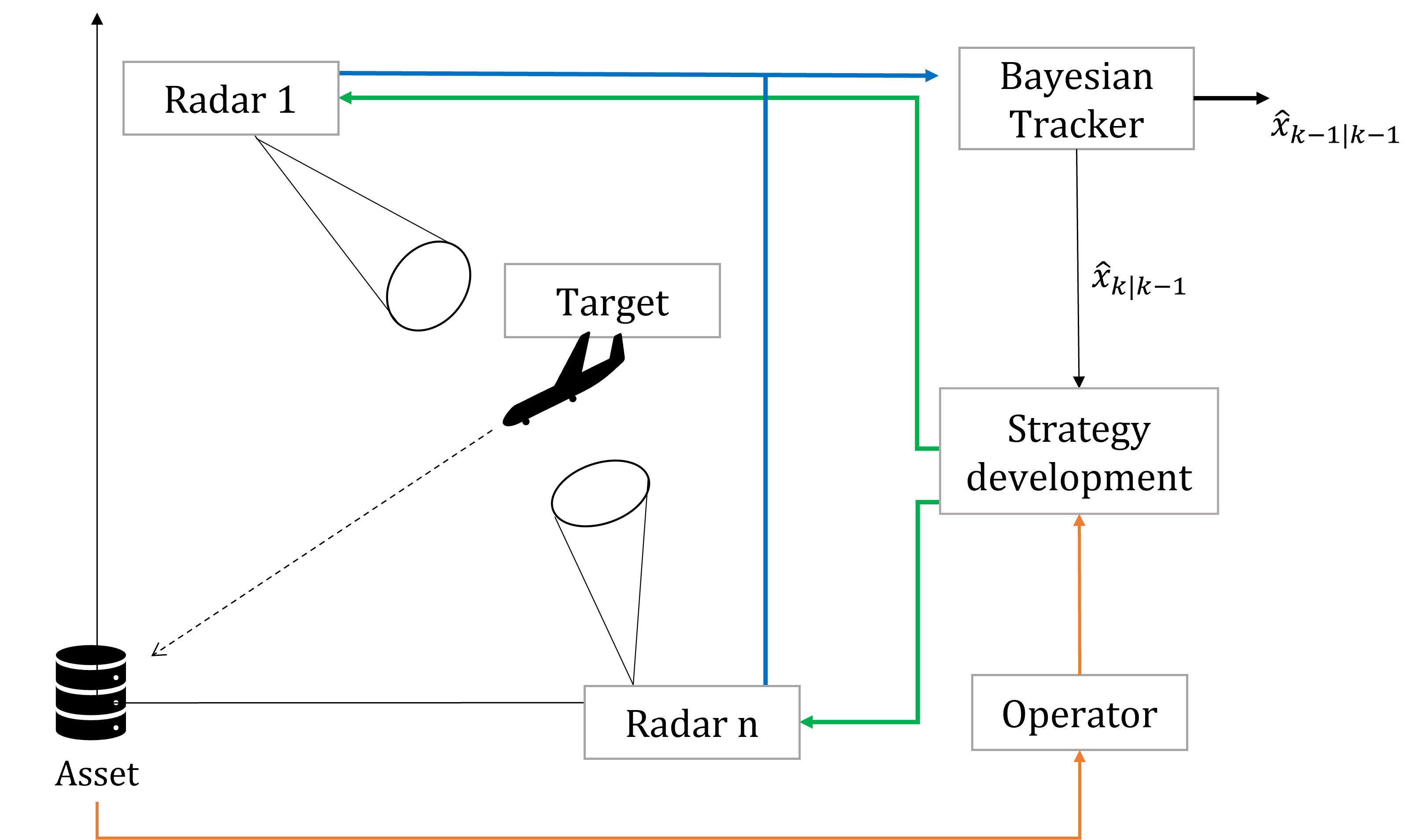}}
\caption{Overview of System Model}
\label{system_model} 
\end{figure}

\section{Threat-based approach framework in cognitive radar}

\subsection{System model}
Fig.~\ref{system_model} represents a military combat situation that is considered in this study. The system model has the following characteristics.
\begin{enumerate}
\item N geographically distributed stationary monostatic radars are deployed in a two-dimensional plane. 
\item Each radar has the same peak transmit power, which is fixed.
\item A single moving target exists.
\end{enumerate}

The position of the \(i\)-th radar, \(p_i^R\), and the target, \(p_k\), at time \(t_k\) can be denoted as follows:
\begin{equation}p_i^R = [x_i^R y_i^R]^T\label{p_i}\end{equation}\begin{equation}p_k = [x_k y_k ] ^T\label{p_k}\end{equation}

The state transition matrix of the target can be represented as follows:
\begin{equation}	X_k=f(X_{k-1})+ V_{k-1} \label{x_k}\end{equation}
where the target state \(X_k = [ x_k \ \dot{x}_k \ y_k \ \dot{y}_k ]^T\) denotes the position and velocity of the target and \(f(\cdot)\) is a model function. \(V_{k-1}\) is process noise, which has a Gaussian distribution with zero mean and covariance \(Q_{k-1}\).

 Note that the target measurement of the \(i\)-th radar at time \(t_k\) can be described using~\eqref{p_i} and~\eqref{p_k}.
\begin{align}
z_{i,k} &= [R_{i,k} \ \varphi_{i,k}]^T + w_{i,k}(T_{i,k}^d)\notag \\ 
&= \left[\lVert p_k - p_i^R \rVert \ \arctan {y_k - y_i^R\over x_k - x_i^R }\right]^T + w_{i,k}(T_{i,k}^d)\label{z_i,k}\end{align}
where \(\lVert *\rVert\) is the Euclidian norm of the function, \(R_{i,k}\) is the range between the radar and the target, \(\varphi_{i,k}\) is the bearing angle between the radar and the target, and $T_{i,k}^d$ is the dwell time of radar \(i\) at $t_k$. The measurement noise \(w_{i,k}\) can be defined as follows: 

\begin{equation}w_{i,k}(T_{i,k}^d) \sim N(0, \ \Gamma_{i,k}) \label{w_i,k}\end{equation}
\begin{equation}\Gamma_{i,k} = diag \{(\sigma_{i,k}^R)^2, (\sigma_{i,k}^\varphi)^2 \} \label{Gamma_i,k}\end{equation}
\(N(0, \Gamma_{i,k})\) is the normal distribution, which has zero mean and covariance \(\Gamma_{i,k}\), and \(diag\{A\}\) denotes a diagonal matrix with A as the value of the nonzero cells.

\begin{figure}[htb!]%
\begin{center}
\centerline{\includegraphics[width=\textwidth]{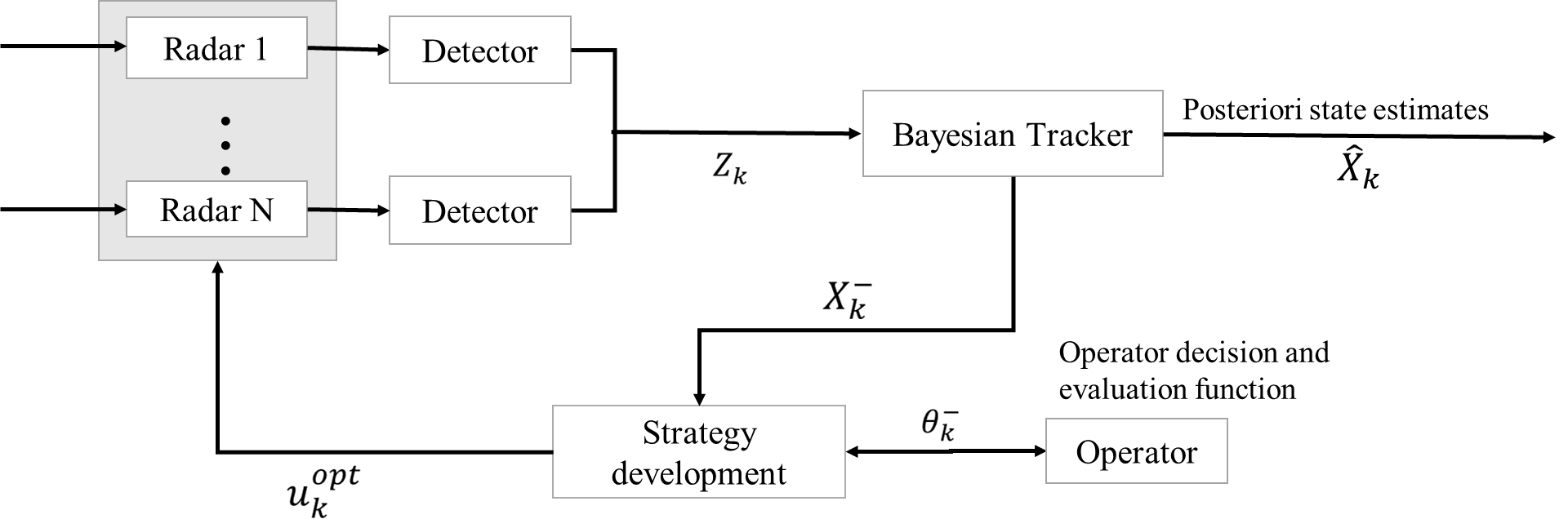}}
\caption{Cognitive action cycle of a radar network.}
\label{Cognitive action cycle}
\end{center}
\end{figure}

In \eqref{Gamma_i,k}, \(\sigma_{i,k}^R\) and \(\sigma_{i,k}^\varphi\) are the theoretical measurement accuracy of the \(i\)-th radar, which can be represented by the following equations~\cite{b24}: 
\begin{equation}\sigma_{i,k}^R ={c\over\left(2\beta_i\sqrt{{SNR}_{i,k}}\right)} \ \propto \left(\beta_i \sqrt{T_{i,k}^d\bar{\sigma_i }/R_{i,k}^4\ }\right)^{-1}\label{sigma_R}\end{equation}
\begin{equation}\sigma_{i,k}^\varphi ={0.628\theta_i^{3dB}\over \sqrt{{SNR}_{i,k}}} \ \propto \left(\sqrt{T_{i,k}^d\bar{\sigma_i }/R_{i,k}^4\ } / \theta_i^{3dB} \right)^{-1}\label{sigma_pi}\end{equation}
where c is the speed of light, \(\beta_i\) is the effective bandwidth of the signal transmitted by the \(i\)-th radar, \(\theta_i^{3dB}\) is the 3 dB beam width of the \(i\)-th radar antenna pattern, \(\bar{\sigma_i }\) is the average target Radar Cross Section (RCS), and \(SNR_{i,k}\) is the signal-to-noise ratio, which is proportional to \(T_{i,k}^d\bar{\sigma_i }/R_{i,k}^4\).

Using \eqref{p_i} - \eqref{sigma_pi}, the measurement equation can be given as 
\begin{equation}	Z_{k}=h(X_{k})+ W_{k}(u_k) \label{z_ik}\end{equation}where 	\(Z_{k}=[z_{1,k}^T  \cdots z_{N,k}^T ]^T\) is the set of measurements of the radar network, \(u_k = [T_{1,k}^d \cdots T_{N,k}^d]^T\) is the radar-target dwell time allocation at time \(t_k\), and
\(W_k (u_k)=[w_{1,k}^T (T_{1,k}^d) \cdots,w_{N,k}^T (T_{N,k}^d)]^T\) is the measurement noise, which has a Gaussian distribution with zero mean and covariance \( R_{u_k}=blkdiag \{ \Gamma_{1,k}, \cdots, \Gamma_{N,k} \}\).
 
\begin{figure}[!t]
\centerline{\includegraphics[width=30pc]
{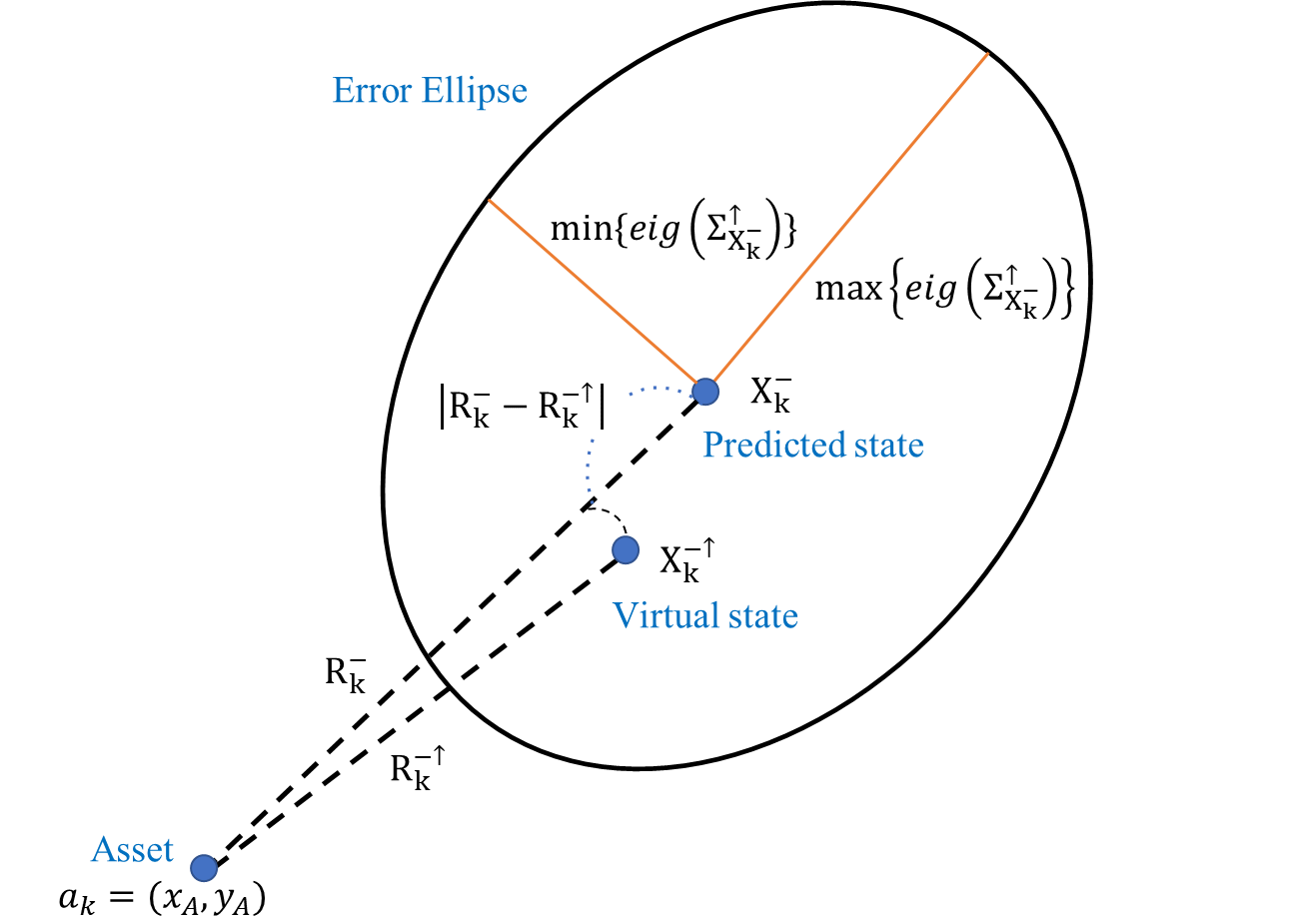}}
\caption{Asset protection scenario.}
\label{fig5}
\end{figure}   

\subsection{Threat model}

The threat function \(\Phi(*)\) considering operational requirements can be constructed as follows: 
\begin{equation}	\theta_k = \Phi(X_k , a_k) \label{theta_k}\end{equation}
where \(\theta_k\) is the threat of the target at time \(t_k\) and \(a_k\) is a variable vector that describes the surrounding battlefield environment (e.g., positions of assets, weights between kinetic parameters). 

To update the state of the target by learning the surrounding environment, the radar network can be modeled as a cognitive action cycle, as shown in Fig.~\ref{Cognitive action cycle}.
According to the cognitive action cycle, high performance can be expected for the next measurement by pre-calculating the optimal solution of the predicted measurement. \( Z_k^\uparrow\), the predicted information for \(Z_k\) considering the previous knowledge before \( t_{k} \), can be defined as
\begin{equation} Z_k^\uparrow \triangleq h(X_k^- )+ W_{k}(u_k) , u_k \in U \label{z_k^uparrow}\end{equation}
where \(U\) is the candidate strategy set. \(X_k^- \sim N(\hat{X}_{k|k-1}, P_{k|k-1})\) is the predicted state of \(X_k\), where $\hat{X}_{k|k-1}$ and $P_{k|k-1}$ are the estimated state and covariance of the target at \(t_{k-1}\)~ \cite{b26}.
Inspired by the Bayesian rule in \eqref{p(x|z)}, the virtual state update $X_k^{- \uparrow}$ can be expressed using $p(X_k^-|Z_k^\uparrow)$~\cite{b2}. 
\begin{equation} p(X_k^-|Z_k^\uparrow) \propto p(X_k^-,Z_k^\uparrow) = p(Z_k^\uparrow|X_k^-)p(X_k^-) \label{p(x|z)}\end{equation}
Since $X_k^{- \uparrow}$ is a Bayesian state estimation problem, its analytical solution is difficult to calculate. Instead, the estimated theoretical bound of PC-CRLB from~\cite{b2} was utilized in this paper.
Subsequently, the virtual threat \( \theta_k^{- \uparrow}\) can also be derived using \( X_k^{- \uparrow}\) and \eqref{theta_k}.
\begin{equation} \theta_k^{- \uparrow} = \Phi(X_k^{- \uparrow} , a_k) 
\label{theta_k^-/uparrow}
\end{equation}
Since \(\theta_k^{- \uparrow}\) is the prediction of \(\theta_k^{-}=\Phi(X_k^{- } , a_k)\), its uncertainty \(\Sigma_{\theta_k^-}^{\uparrow}\) can be represented using the definition of the predicted conditional mean square error (PC-MSE)~\cite{b21}.
\begin{equation}\Sigma_{\theta_k^-}^{\uparrow}=E_{X_k^-,Z_k^\uparrow} [(\theta_k^- - \theta_k^{- \uparrow})^2] \label{Sigma_k^-,uparrow} \end{equation} 
where \(E_x[*]\) is expectation with respect to x. 
\subsection{Restriction of the threat function}
As a surveillance system, radar must provide the operator with good threat accuracy so that the operator can properly recognize the battlefield situation and respond accordingly. Since \(\theta_k^{- \uparrow}\) is a Bayesian estimation problem \cite{b27}, the efficiency can be expressed using the variance of the estimator expressed in \eqref{Sigma_k^-,uparrow}. The method of calculating the variance may be different depending on the definition of \(\Phi(X_k , a_k)\) in \eqref{theta_k} with respect to the considered environment. In this study, threat function restrictions are derived considering a scenario in which the radar system protects fixed combat assets (e.g., operational command posts) from incoming missiles, as shown in Fig.~\ref{fig5}.

The distance \(R\) between the asset and the target can be defined using the given asset position \(a_k=(x_A, y_A)\), where $x_A$ is the x-position and $y_A$ is the y-position of the asset, and the target state \(X_k\):

\RestyleAlgo{ruled}
\SetKwComment{Comment}{/* }{ */}
\begin{algorithm}[hbt!]
\caption{Target-tracking process in the cognitive radar framework}\label{alg_ov}
\textbf{Initialization: }\\
\(\quad t_k=1;\)\\
\(\quad \hat{X}_{0|0} = X_0;\)\\
\(\quad P_{0|0} = P_0\)\\ 
\While{ $t_k < t_{k_{max}} $}{ 
\textbf{Step 1. Estimation}\\
\(\hat{X}_{k|k-1} = f(\hat{X}_{k-1|k-1})\); 
\(A_k\) = Jacobian of \(f\) at \(t_k\)\;
\(P_{k|k-1}\) = \(A_k P_{k-1|k-1} A_k^T + Q_{k-1}\)\; 
\textbf{Step 2. Decision making}\\
\(X_k^- \sim N(\hat{X}_{k|k-1}, P_{k|k-1})\)\;
Calculate \(\max\{{eig(\Sigma_{X_k^-})}\}\)\;
Formulate the optimization problem of \eqref{optimal_1}\;
Obtain \(u_k\) by solving the optimization problem\;
\textbf{Step 3. State Update}\\
	\(\hat{X}_{k|k} = \hat{X}_{k|k-1}\)\;	\For{\(i = 1 \)\KwTo $i_{max}$}{ 
  	\If{\(u_{i,k} >0\)}{
    \(Z_{i,k}=h(X_{i,k})+ W_{k}(u_k)\)\; 
    \(H^i_k\) = Jacobian of \(h\) at \(\hat{X}_{k|k-1}\)\;
    Calculate the Kalman Gain \(K_k\)\;
\(\hat{X}_{k|k} = \hat{X}_{k|k-1} + K_k(Z_{i,k} - h(\hat{X}_{k|k-1}))\)\;
    \(P_{k|k} = P_{k|k-1} - K_kH^i_kP_{k|k-1}\)\;
    }
	}
$t_k=t_{k}+1$;}
\end{algorithm}

\begin{equation}a_k=(x_A, y_A )\label{a_k} \end{equation}
\begin{equation}R(a_k,X_k)=\sqrt{(x_A-x_k )^2+(y_A-y_k )^2 }\label{R(a,X)} \end{equation}
The distance between the target and the asset can be considered a parameter for calculating the threat. 
Therefore, \(\theta_k\) can be expressed using \eqref{R(a,X)}, and \(g\) is the formula for finding the threat \(\theta\) using \eqref{R(a,X)}: 
\begin{equation}\Phi(X_k, a_k) = g(R(a_k,X_k))= \theta_k \label{Phi(R)} \end{equation}
Because R is the distance, \(g(x)\) is defined only when \(x \ge 0\), and we define the threat as a positive value. Since the asset needs to be protected, the threat becomes smaller as the distance between the asset and the target increases. Therefore, the corresponding restriction is applied, where \(\Delta\) denotes the instantaneous rate of change of the amount.
\begin{equation}\Delta g \Delta x \le 0\label{restrictionii} \end{equation}
Using \eqref{Sigma_k^-,uparrow} and \eqref{Phi(R)}, $\Sigma_{\theta_k^-}^{\uparrow}$ can be expressed using \(g(x)\) and R.
\begin{equation}\ \Sigma_{\theta_k^-}^{\uparrow}=E_{X_k^-,Z_k^\uparrow} [\{g(R_k^-)-g(R_k^{- \uparrow})\}^2] \label{MSE_theta} \end{equation}
where \(R_k^{- \uparrow} = R(a_k, X_k^{- \uparrow}\) ) and \(R_k^- = R(a_k, X_k^-)\).

Since \(g(*)\) is used to find the bound value of \(R_k^{- \uparrow}\), it is difficult to generalize \eqref{MSE_theta} unless \(g(x)\) is specifically given. If \(g(x)\) instead has a constraint, \eqref{MSE_theta} can be obtained by using a previously known performance measure of target state estimation such as the eigenvalue~\cite{b26}.

Let us assume that there exists a function \(f\) that has the following identity relationship with g, where \(R^+\) is non negative real number, \(x_R, y_R \in R^+ \).
\begin{equation}\ f((x_R -y_R)^2)=\{g(x_R)-g(y_R)\}^2 \label{proof1} \end{equation}

\textit{Proposition 1} : If $a<b$, then $f(a^2) < f(b^2)$. 

\textit{Proof}: When $a<b$, $0<g(b)<g(a)<g(0)$ holds since $\theta_k $ is positive, and \eqref{restrictionii}. $\{g(0)-g(a)\}^2 < \{g(0) -g(b)\}^2$ then also holds. Using \eqref{proof1}, $f(a^2) < f(b^2)$ holds.

Note that if an arbitrary pair \(\alpha, \beta \in R^+ \) satisfies \eqref{proof1} and \(\alpha \ge \beta\), it can be expressed as follows.
\begin{equation}\ f((\alpha - \beta)^2)=\{g(\alpha)-g(\beta)\}^2 \label{proof4} \end{equation}

Additionally, we can substitute \(0\) for \(x_R\) and \((\alpha-\beta)\) for \(y_R\) in \eqref{proof1}.
\begin{align}\ f((\alpha-\beta)^2)&=\{g(\alpha-\beta)-g(0)\}^2
\label{proof2}\end{align}
Using \eqref{proof4} and \eqref{proof2}, \eqref{proof3} can be shown. 
\begin{align}\ \{g(\alpha-\beta)-g(0)\}^2 = (g(\alpha) - g(\beta))^2 \label{proof3} 
\end{align}
Since \(g(\alpha-\beta) < g(0)\), and we have \(g(\alpha) < g(\beta)\) by \eqref{restrictionii}, we can obtain the restriction on \(g(*)\) that includes a pair \(\alpha, \beta\).
\begin{equation}
g(\alpha-\beta) - g(\alpha) + g(\beta) - g(0) = 0    
(\alpha \ge \beta \ge 0)
\label{condition1}\end{equation}
If the same process is repeated in the case of \(\alpha <\beta\), the following condition is obtained.
\begin{equation}
g(\beta-\alpha) - g(\beta) + g(\alpha) - g(0) = 0    
(\beta \ge \alpha \ge 0)
\label{condition2}\end{equation}
Since the asset position is fixed, \(|R_k^{- \uparrow} - R_k^-|\) always has a smaller value than the major axis of the position error in the ellipse of \(X_k^{- \uparrow}\) in Fig.~\ref{fig5}. That is, \(|R_k^{- \uparrow} - R_k^-|\), which can be considered the performance of the estimated threat accuracy, has as an upper bound the major axis of the error ellipse of \(X_k^{- \uparrow}\). 
The error ellipse of \(X_k^{- \uparrow}\) is a geometrical expression of the PC-MSE matrix of \(X_k^{- \uparrow} \), which can be expressed as follows: 
\begin{equation}
\Sigma_{X_k ^-}^{\uparrow}= E_{X_k^-,Z_k^\uparrow} [(X_k ^- - X_k ^{- \uparrow})(X_k ^- - X_k ^{- \uparrow})^T] 
\label{Sigma_Xk^-,uparrow}
\end{equation}

However, usually $\Sigma_{X_k^-}^{\uparrow}$ cannot be expressed analytically, so the PC-CRLB of \(\Sigma_{X_k^-}^{\uparrow}\), \(B^{- \uparrow}_k\), can be obtained instead. \(B^{- \uparrow}_k\) is the smallest error ellipse of \(X_k^{- \uparrow}\), and its main axis is the maximum eigenvalue of \(B^{- \uparrow}_k\) (See Appendix I). This can be expressed as follows: 
\begin{equation}
 E_{X_k^-,Z_k^\uparrow}[|R_k^- - R_k^{- \uparrow}|] \le \max\{eig(B^{- \uparrow}_k)\}
\label{R_k - R_k^-}\end{equation}
where \({eig(*)}\) denotes all eigenvalues of the function and \(max(*)\) is the maximum value of the elements. 

Then, combining the above \textit{Proposition~1}, \eqref{MSE_theta}, \eqref{proof1}, and \eqref{R_k - R_k^-} yields
\begin{align}E_{X_k^-,Z_k^\uparrow}[ (\theta_k^- - \theta_k^{- \uparrow})^2] &=E_{X_k^-,Z_k^\uparrow}[\{g(R_k^-)-g(R_k^{- \uparrow})\}^2] \notag 
\\&= E_{X_k^-,Z_k^\uparrow}[\ f((R_k^- - R_k^{- \uparrow})^2) ]\notag 
\\&=f(E_{X_k^-,Z_k^\uparrow}[\ (R_k^- - R_k^{- \uparrow})^2 ])\notag
\\&\le f(\max\{eig(B^{- \uparrow}_k)\}^2)
\label{combined theta_k}\end{align}
In \eqref{combined theta_k}, $f(\max\{eig(B^{- \uparrow}_k)\}^2)$ is the maximum value of the threat MSE $\theta_k^-$. The maximum error is the largest difference the estimator can have, so it can be viewed as a performance indicator. Therefore, the problem of finding the accuracy of the threat can be replaced by the problem of finding the accuracy of the state generally in cognitive radar. In other words, in general cases, for a given $g(*)$ that satisfies \eqref{condition1} or \eqref{condition2}, $f(*)$ exists and can be used to determine the threat accuracy using the eigenvalue of the state estimate. The detailed procedure will be explained in Section IV by using the specific function $g(*)$.

\subsection{Optimization formulation and dwell time allocation}
 Since the target state estimation problem is a Bayesian estimation problem, the efficiency can be expressed using its covariance. The PC-CRLB supplemented with the Bayesian Cramer--Rao Lower Bound (BCRLB) is a method commonly used to calculate the lower bound of covariance, and it is utilized in this study for performance evaluation in cognitive radar. 
 In a situation where dwell time is a radar resource, the cost function \(C(u_k)\) of the resource allocation output \(u_k\) can be expressed as \(C(u_k) = 1^Tu_k\). If the constraint of the optimization problem is formulated to keep the threat accuracy from reaching a certain lower limit \(P\), it can be expressed as follows.
 \begin{align}
	&\\argmin_{u_k} C(u_k) \notag \\
\text{subject to}
	&\ 0 \le u_k \le T_k^{d,max} \notag \\
 	&\ M(\theta_k^{- \uparrow} ; u_k) > P  
\label{optimal_1}
\end{align}

 where \(M(\theta_k^{- \uparrow} ; u_k)\) is a function that evaluates \(\theta_k^{-\uparrow}\).
 M can be defined using known performance measures of the covariance matrix such as eigenvalue or trace. The optimal allocation strategy, \(u_k^{opt}\), can be achieved by solving optimization problem \eqref{optimal_1}. \(u_k^{opt}\) can be applied to a radar network by controlling the measurement noise in \eqref{z_i,k}. 

\section{Example of the threat-based approach}
Let us assume that \(g(*)\) has the following form, which satisfies \eqref{restrictionii} and \eqref{condition1}.
\begin{equation}
g(R_k) = aR_k + b
\label{g(R_k)}
\end{equation}
where \(a < 0\), \(b > 0\) , and \(R_k\) is \(R(a_k,X_k)\) in \eqref{R(a,X)}.
Using the definition of $f(*)$ in \eqref{proof1}, $f(x) = a^2x$ can be obtained.
Substituting \eqref{g(R_k)} into \eqref{combined theta_k} yields the following.

\begin{align}
E_{X_k^-,Z_k^\uparrow}[ (\theta_k^- - \theta_k^{- \uparrow})^2] 
&=f(E_{X_k^-,Z_k^\uparrow}[\ (R_k^- - R_k^{- \uparrow})^2 ])\notag\\
&=a^2E_{X_k^-,Z_k^\uparrow}[(R_k^- - R_k^{- \uparrow})^2] \notag \\
&\le a^2\max\{eig(B^{- \uparrow}_k)\}^2
 \label{example2} \end{align}

Since $a^2\max\{eig(B^{- \uparrow}_k)\}^2$ is proportional to the absolute value of $\max\{{eig(B^{- \uparrow}_k)}\}$ and \(\max\{{eig(B^{- \uparrow}_k)}\}\) is positive, \(M(\theta_k^{- \uparrow}; u_k)\) in \eqref{optimal_1} can be considered to be \(\max\{{eig(B^{- \uparrow}_k)\}}\). \eqref{optimal_1} can be transformed into an SOCP~\cite{b21}. The problem can then be solved using well known cone-programming optimization solvers.

Algorithm~\ref{alg_ov} explains the full flow of the system, where \(k_{max}\) is the maximum time frame, \(i_{max}\) is the number of radars, and the Extended Kalman Filter (EKF) is used as a Bayesian tracker.

\begin{figure*}%
\begin{center}
\subfloat[Scenario 1] {\includegraphics[width=0.5\textwidth]{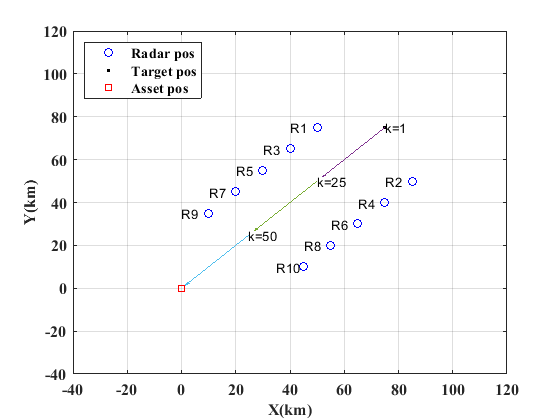}}\subfloat[Scenario2]{\includegraphics[width=0.5\textwidth]{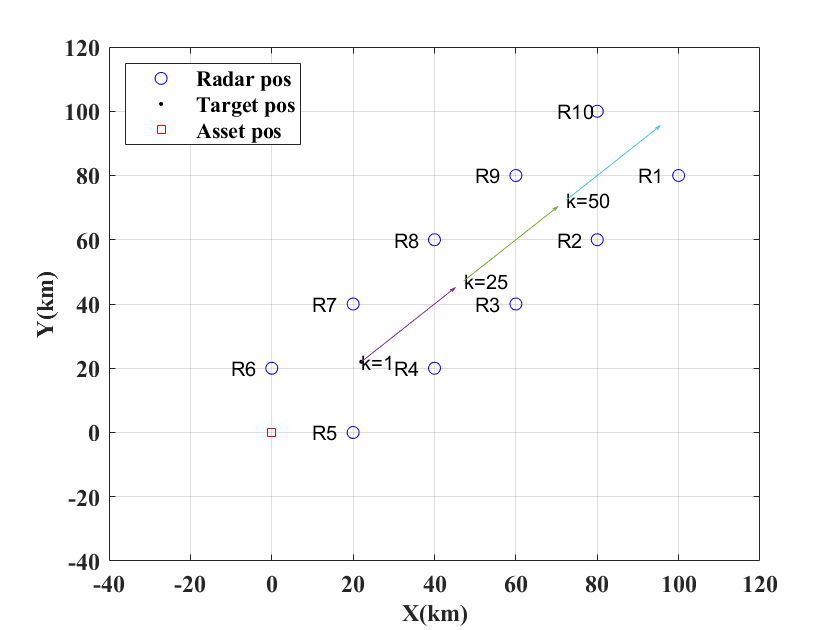}}
\end{center}
\caption{Two scenarios of radar--target--asset deployment. (a): Gathered radars approaching assets. (b): Scattered radars moving away from assets. }
\label{radar_deployment}
\end{figure*}

\begin{figure*}%
\begin{center}
\subfloat[Time consumption]{{\includegraphics[width=0.5\textwidth]{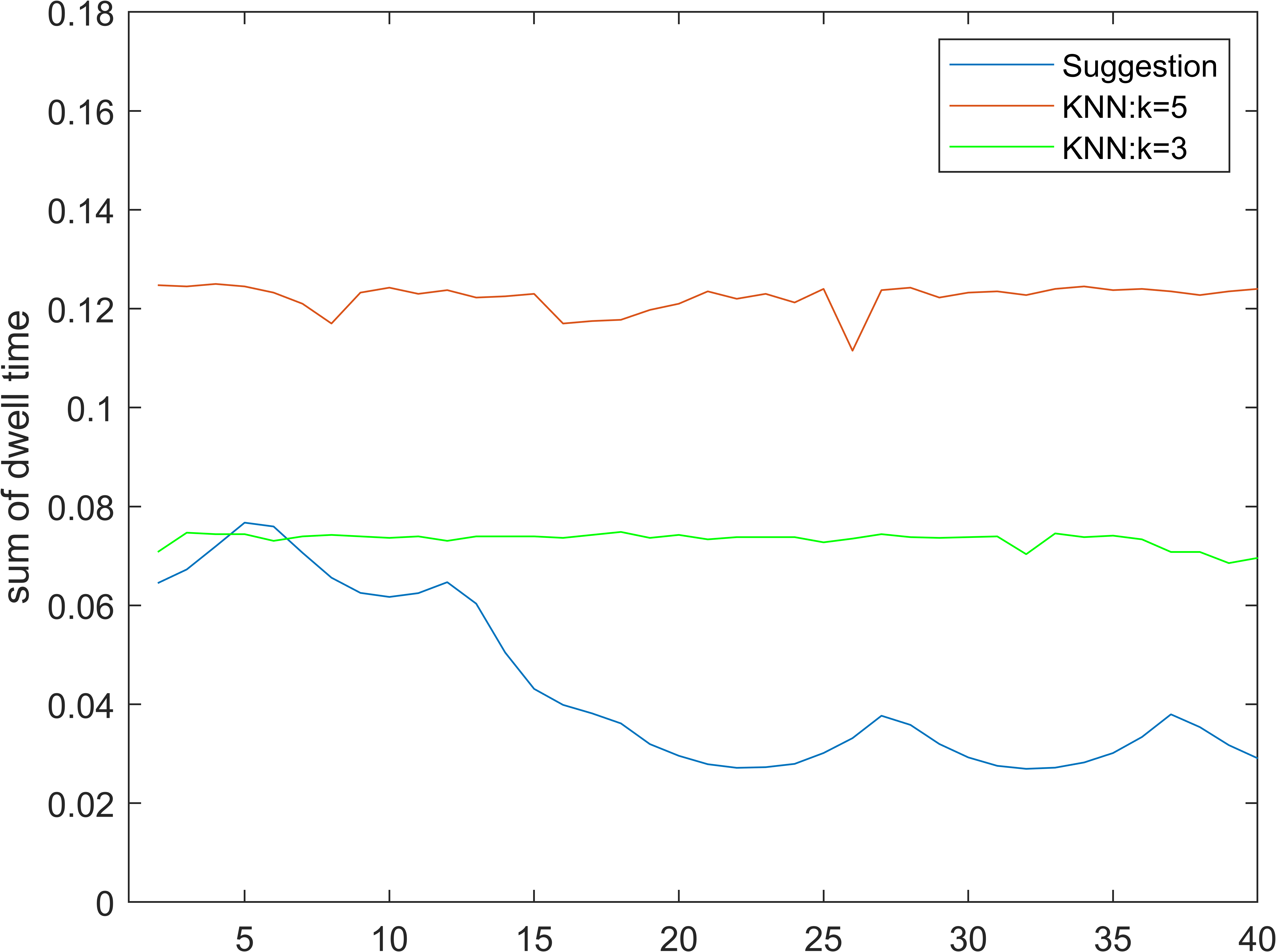} }}%
\subfloat[Time allocation to each radar]{{\includegraphics[width=0.5\textwidth]{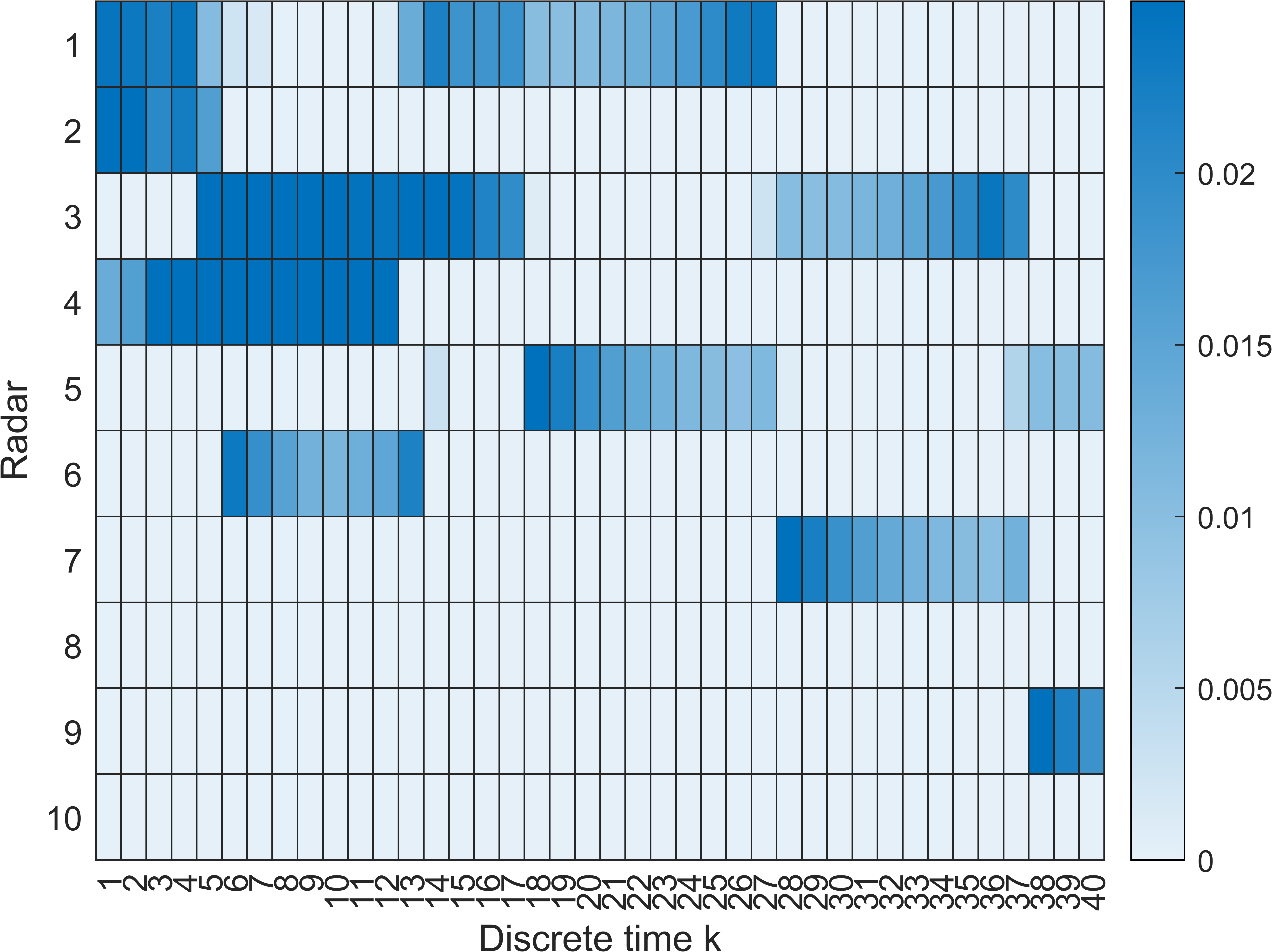}}}%
\hfill
\subfloat[Threat difference]{{\includegraphics[width=0.5\textwidth]{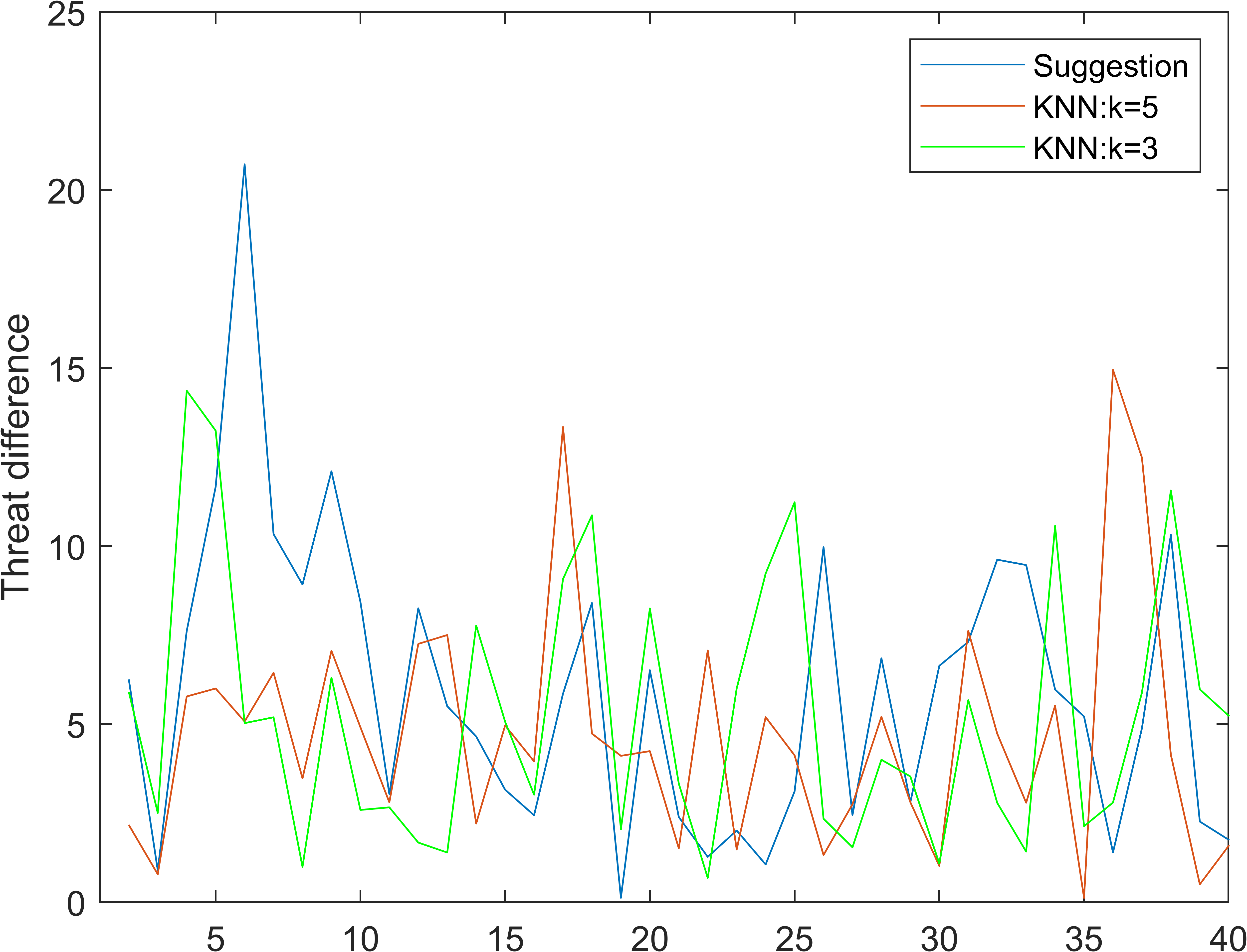} }}%
\subfloat[Product of time consumption and threat difference]{{\includegraphics[width=0.5\textwidth]{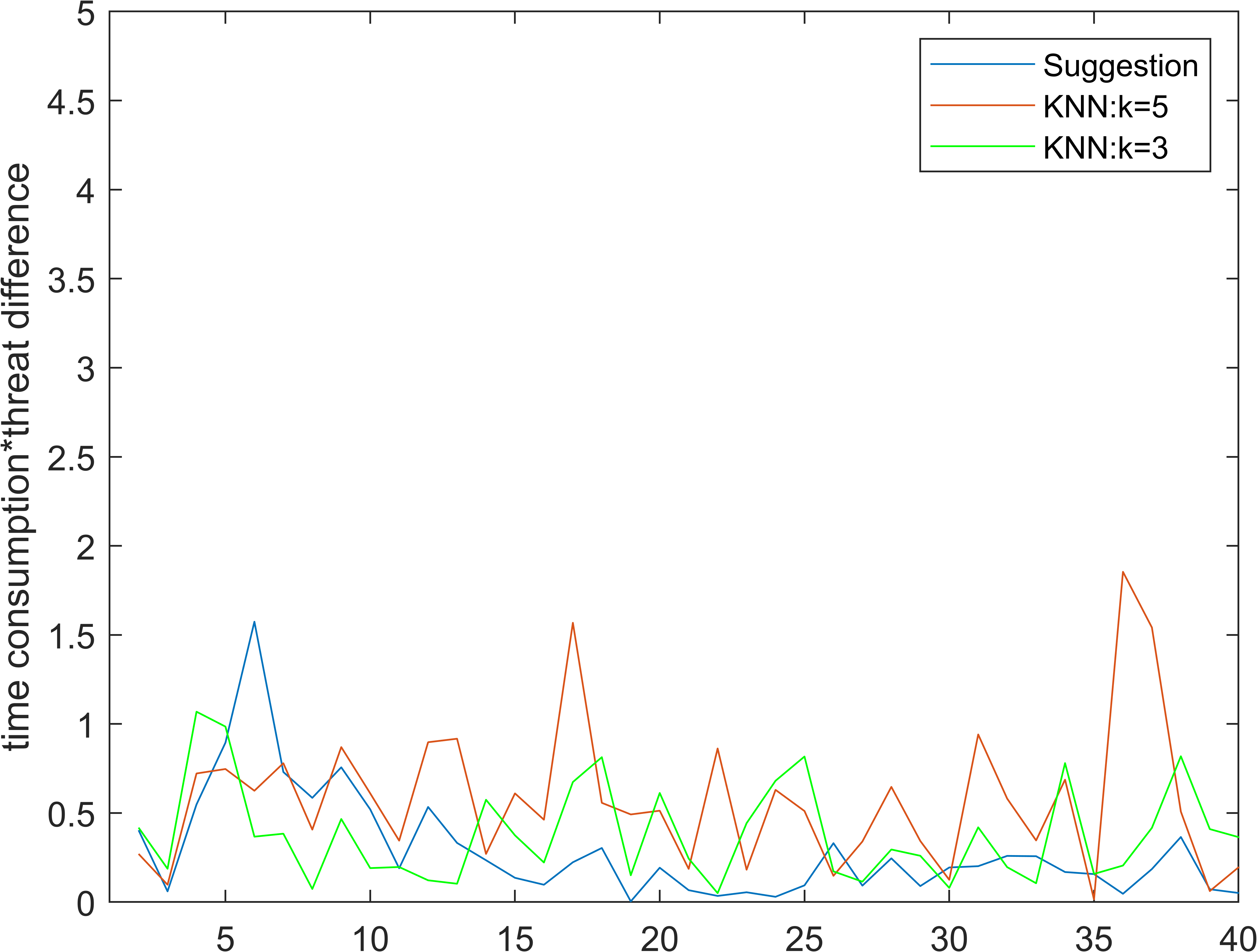} }}%
\end{center}
\caption{Simulation results of Scenario 1 }
\label{result1}%
\end{figure*}

\begin{figure*}%
\begin{center}
\subfloat[Time consumption]{{\includegraphics[width=0.5\textwidth]{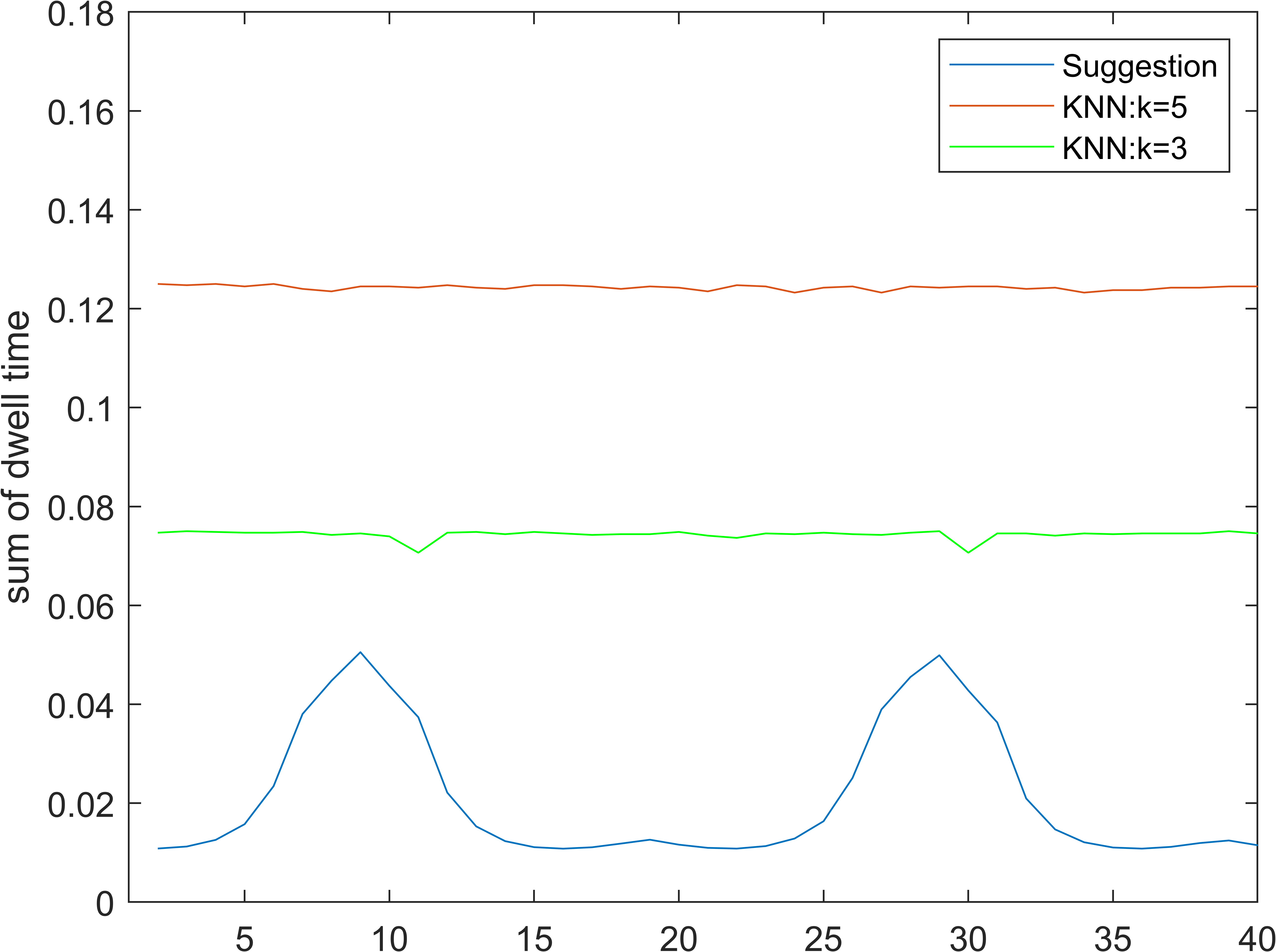} }}%
\subfloat[Time allocation to each radar]{{\includegraphics[width=0.5\textwidth]{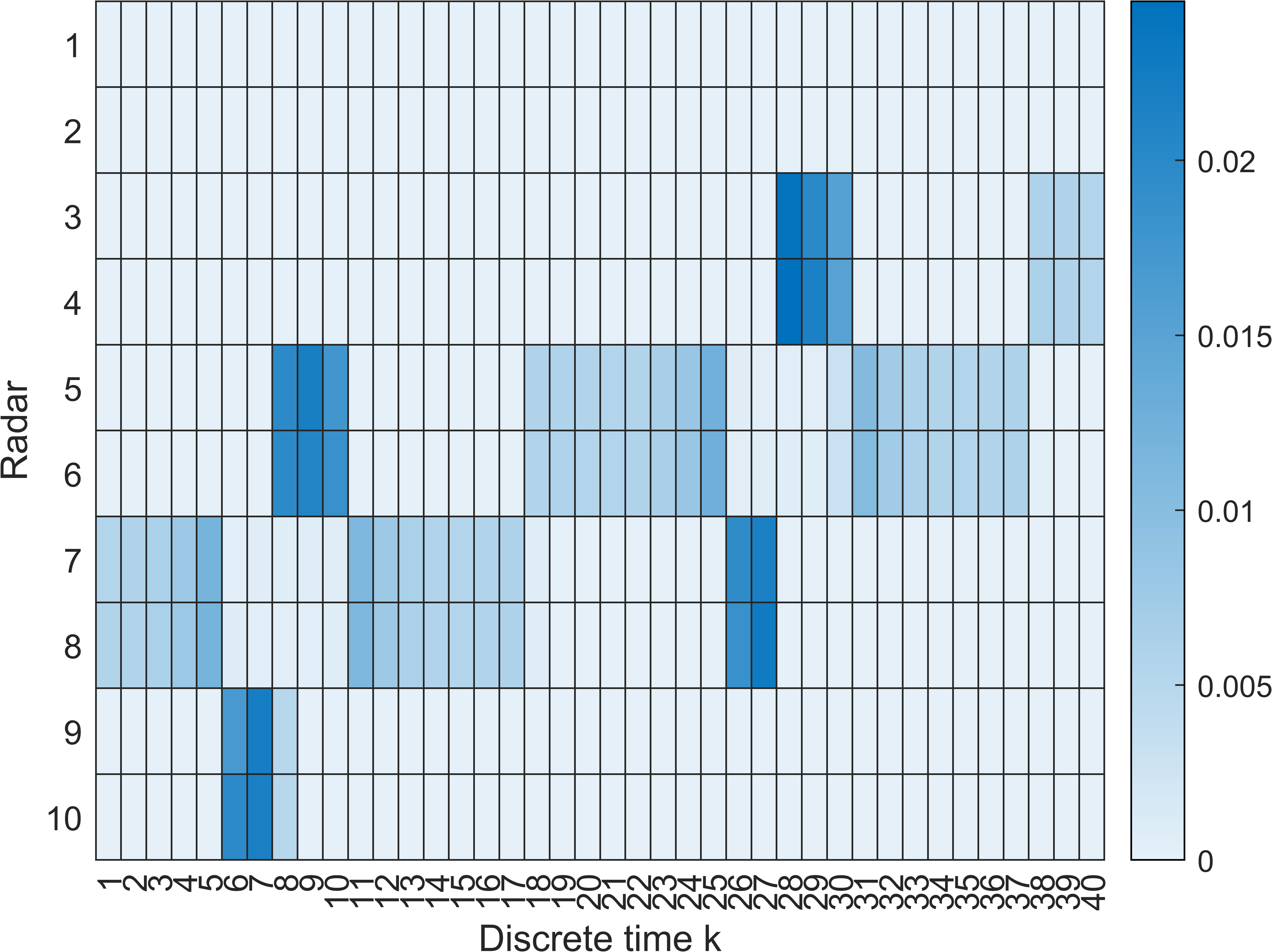}}}%
\hfill
\subfloat[Threat difference]{{\includegraphics[width=0.5\textwidth]{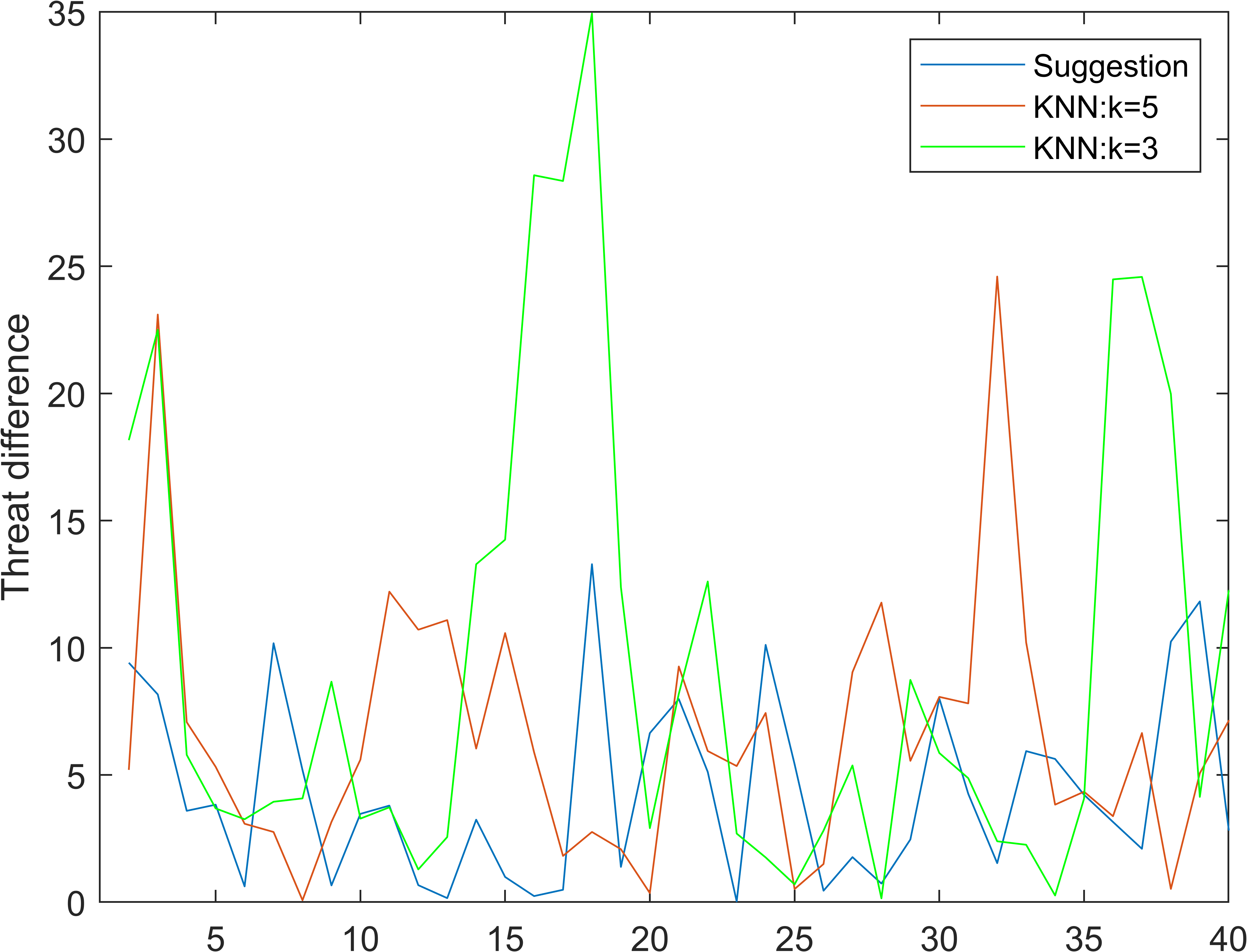} }}%
\subfloat[Product of time consumption and threat difference]{{\includegraphics[width=0.5\textwidth]{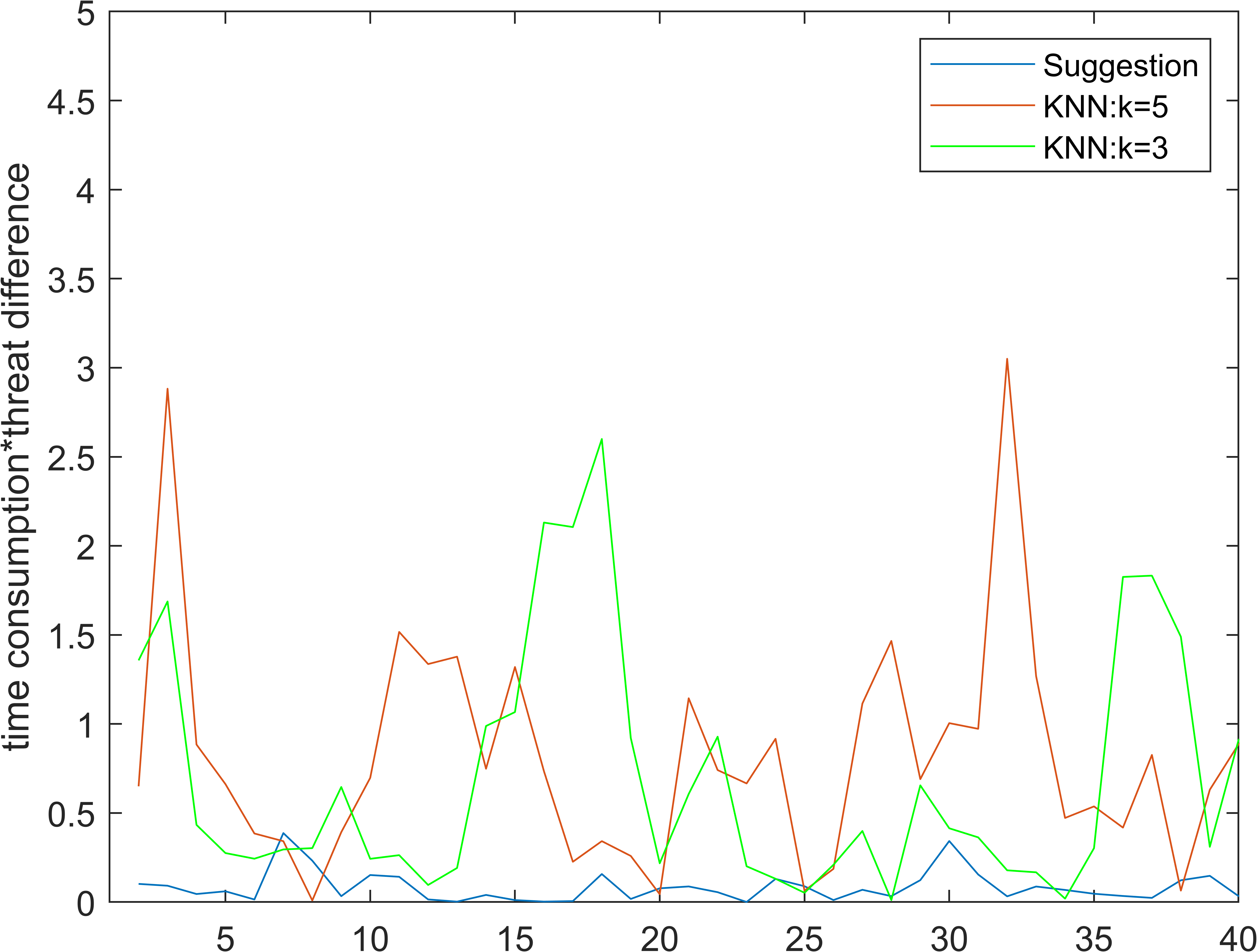} }}%
\end{center}
\caption{Simulation results of Scenario 2 }
\label{result2}%
\end{figure*}

\section{Simulation results}
A numerical simulation was conducted to assess the performance of the proposed algorithm. A system of 10 monostatic vehicle mounted radars was considered. Since each radar was mounted on a vehicle, there was a limited dwell time, and efficient resource allocation was important in maximizing the capacity of the radar. Simulations were conducted on two different scenarios to prove the robustness of the proposed algorithm. (See Fig. \ref{radar_deployment}.)
We assumed all radars to have the same invariant effective bandwidth and average RCS: \(\beta_i \)= 1/3 MHz, \(\theta^{3dB}_i\) = 10 mrad, and \(\bar{\sigma}_{i|k:1:k-1}\)= 1\(m^2\). The time interval was 25 ms, and the maximum time frame was 50.
The target was tracked using a nearly constant-velocity model of the EKF \cite{b21}. \( Q = \begin{bmatrix} \triangle T^3/3 & \triangle T^2/2 \\ \triangle T^2/2 & \triangle T \end{bmatrix} \otimes \kappa I_2\), where \(I_2\) is the \(2 \times 2\) identity matrix, \(\otimes\) is the Kronecker product, \(\kappa\) is \(1m^2/s^3\), and \(P_0^- = 250^2diag \{1,1,2/\triangle T^2, 2/\triangle T^2\} m^2\). Local targets from multiple radars were associated as a single system target using a common method~\cite{b25}. The threat function used was $\theta_k = -100R_k+100$. The radars were considered stationary, and the target moved at 142.4 m/s.

The results are shown in Fig.~\ref{result1} and Fig.~\ref{result2}.
The proposed algorithm was compared with the KNN algorithm, which allocates limited time resources to the nearest k radars. The performance was compared with the total time allocated to each radar when they had the same average threat MSE value, as shown in Fig.~\ref{result1}(c) and Fig.~\ref{result2}(c). Fig.~\ref{result1}(a) shows the performance comparison between the proposed algorithm and the KNN algorithm in Scenario 1. It shows that the proposed algorithm produces the same average threat error with less time. For more detail, Fig.~\ref{result1}(b) shows a heatmap of the proposed algorithm in Scenario 1. It indicates that the proposed algorithm flexibly adapts to changes in the external environment by allocating time resources to various radars. This can be clearly seen: time is allocated to four radars at $t_k$=5; three radars at $t_k$=1~4, 6~14, 27, and 37; and two radars at $t_k$=15~26, 28~36, and 38~40 in Fig.~\ref{result1}(b). Additionally, it can be seen that the algorithm allocates more time as the target and radar become closer to increase the threat accuracy. Fig.~\ref{result1}(d) shows the product of the sum of allocated time in Fig.~\ref{result1}(a) and the threat MSE in Fig.~\ref{result1}(c) to show a more clear result. In general, since the sum of the allocated time and threat accuracy are traded off, the product of these two values can be used to compare the two performance indicators at the same time. The results in Fig.~\ref{result1}(d) show that the proposed algorithm has the best performance in most of the timeframe, especially after $t_k=14$. That is, the proposed algorithm is efficient in most of the timeframe even when the time and threat accuracy are considered at the same time for every frame.

The superiority of the algorithm is more evident in Scenario 2, where the target is distributed. Fig.~\ref{result2}(a) shows that the proposed algorithm uses fewer time resources at all times in Scenario 2. In addition, Fig.~\ref{result2}(d), which considers the threat accuracy at the same time, shows the best performance in almost all timeframes. The results in these two different scenarios indicate that the proposed algorithm robustly adapts to changes in the external environment.

\section{Conclusion} 
Previous research on cognitive radar mainly aimed to improve the accuracy of the target position. In this paper, a threat-based approach is utilized in an air defense target tracking scenario with cognitive radar to allocate tracking resources by calculating the target's threat accuracy. The threat of the target is regarded as the distance between the target and the asset, and the detailed parameters can be set by the radar operator according to the operational environment. In particular, it is noteworthy that the proposed method can bring operational considerations into the decision process of cognitive radar. This is more similar to human behavior than previous methods, and this research can be extended by considering artificial intelligence techniques such as deep learning to train the operational decision making process. The proposed algorithm is proved to achieve better target threat accuracy with similar time resources as other resource allocation methods through numerical simulation.
 
\appendix
\section{Calculating the eigenvalue of the error covariance}
\textit{Proof:} The lower bound of the major axis of \(X_k^- - X_k^{- \uparrow}\) can be obtained using the eigenvalue of the error covariance \cite{b2}. Each eigenvalue represents an axis of the ellipse, and a larger eigenvalue represents a longer axis.
The PC-MSE of \(X_k^- - X_k^{- \uparrow}\) can be expressed as in \eqref{Sigma_Xk^-,uparrow}. 
A bound of \eqref{Sigma_Xk^-,uparrow} \(B^{- \uparrow}_k\) can be expressed as follows : 
\begin{equation}
J_B^\uparrow(p_k^-) = \sum_{k=0}^N T^{d}_{i,k} E_{x_k^-}[J(\varphi_{i,k})] + (A-BC^{-1}B^T)
\label{J_B_1}
\end{equation}
The PC-BIM of \(p^-_k = [X_k^-]_{2\times1}\) can be expressed using \( J_P^\uparrow(X_k^-) =\begin{bmatrix}A&B \\ B^T&C\end{bmatrix} \), \(A, B, C \in \mathbb{R}^{2 \times 2}\) \cite{b29}: 
\begin{equation}
J_B^\uparrow(p_k^-) = \sum_{k=0}^N T^{d}_{i,k} E_{x_k^-}[J(\varphi_{i,k})] + (A-BC^{-1}B^T)
\label{J_B_2}
\end{equation}
\eqref{J_B_2} can be simplified as follows :
\begin{equation}
J_B^\uparrow(p_k^-) = \sum_{k=0}^N T^{d}_{i,k} J_{i,k} 
\end{equation}
where $T^{d}_{0,k} = 1$, $J_{0,k} = A-BC^{-1}B^T$, $J_{i,k} = E_{x_k^-}[J(\varphi_{i,k})]$.  
The equivalent PC-CRLB is
\begin{equation}
B_{p_k^-, k}^{- \uparrow} =\left[ J_B^{\uparrow} (p_k^-) \right]^{-1} = \left[ B_{k}^{- \uparrow}\right]_{2 \times 2}
\end{equation}
Using singular value decomposition (SVD), \(J_{i,k}\) can be written as follows :
\begin{align}
J_{i,k} &= J(\mu_{1,i,k}, \mu_{2,i,k}, \vartheta_{i,k}) \\
&= U_{\vartheta_{i,k}} \begin{bmatrix} \mu_{1,i,k} & 0 \\ 0 & \mu_{2,i,k} \end{bmatrix} U^T_{\vartheta_{i,k}}
\label{J_{i,k}}
\end{align}
where \(\mu_{1,i,k}, \mu_{2,i,k}\) are the eigenvalues of \( J_{i,k}, U_{\vartheta_{i,k}}\) is the rotation matrix with angle \(\vartheta_{i,k}\).
Let \(J_B^\uparrow(p_k^-) = J(\mu_{1,k}, \mu_{2,k}, \vartheta_k)\), according to the relationship \(J_B^\uparrow(p_k^-) = \sum_{k=0}^N T^{d}_{i,k} J(\mu_{1,i,k}, \mu_{2,i,k}, \vartheta_{i,k})\). 
Then, the eigenvalues of \(J_B^\uparrow(p_k^-)\), $\{ 1 / {\mu_{1,k}}, 1 / {\mu_{2,k}} \}$, can be expressed as follows \cite{b2}:
\begin{align}
{1 \over \mu_{1,k}}, {1 \over \mu_{2,k}} = {1 \over 2} & \left[ \sum_{k=0}^N T^{d}_{i,k} (\mu_{1,i,k} + \mu_{2,i,k}) \right. \\  & \left. \pm \begin{Vmatrix} \sum_{k=0}^N T^{d}_{i,k} (\mu_{1,i,k} - \mu_{2,i,k}) u(2\vartheta_{i,k}) \end{Vmatrix} \right ]
\end{align}
Using the function above, we can obtain the eigenvalue of the error covariance.
\begin{align}
eig(B) &= \{ {1 \over \mu_{1,k}}, {1 \over \mu_{2,k}} \}\notag \\
&= \{ {1 \over \mu_{1,k}}, {1 \over \mu_{2,k}} \}
\label{eig_B}
\end{align}

\bibliographystyle{unsrt}

\begin{thebibliography}{40}

\bibitem{b24} M. Richards, W. Holm, and J. Scheer, \emph{Principles of Modern Radar: Basic Principles.} London, UK : SciTech Publishing, 2010. 

\bibitem{b26} C. Yang, L. Kaplan, and E. Blasch, “Performance Measures of Covariance and Information Matrices in Resource Management for Target State Estimation,” \emph{IEEE Transactions on Aerospace and Electronic Systems}, vol.$ 48$, no. $3$, \emph{pp. 2594-2613}, July2012.

\bibitem{b27} H. Trees and K. Bell, \emph{Bayesian Bounds for Parameter Estimation and Nonlinear Filtering/Tracking.} New York, NY, USA : Wiley, 2007.

\bibitem{b1}G. W. Stimson, H. D. Griffiths, C. J. Baker, and D. Adamy, \emph{Stimson’s Introduction to Airborne Radar, in 3}rd ed. Stevenage, UK : SciTech, 2014.

\bibitem{b25} M. Skolnik, \emph{Introduction to Radar Systems, 3}rd ed. New York, NY, USA: McGraw-Hill College, 2003.

\bibitem{b4} Y. Su, T. Cheng, Z. He, and X. Li, “Adaptive simultaneous multibeam resource management for colocated MIMO radar in multiple targets tracking,” \emph{Signal Process.}, vol. $172$, July 2020. 

\bibitem{b15} X. J. Gu, X. M. Wang, K. R. Zhao, and Y. Li, “Autonomous resource management system of netted radar for tactical aircrafts.”, in \emph{2009 IEEE International Conference on Control and Automation}, Christchurch, New Zealand, 2009. 

\bibitem{b16} P. W. Moo, and Z. Ding, “Coordinated radar resource management for networked phased array radars,” \emph{IET Radar, Sonar \& Navigation}, vol. $9.8$, pp. \emph{1009-1020}, 2015.

\bibitem{b17} Q. Han, M. Pan, W. Long, Z. Liang, and C. Shan, “Joint Adaptive Sampling Interval and Power Allocation for Maneuvering Target Tracking in a Multiple Opportunistic Array Radar System,” \emph{Sensors}, vol. $20$, no. $4$, pp.\emph{981}, 2020. 

\bibitem{b18} M. Xie, W. Yi, T. Kirubarajan, and L. Kong, “Joint node selection and power allocation strategy for multitarget tracking in decentralized radar networks,” \emph{IEEE Transactions on Signal Processing}, vol. $66$, no. $3$, pp.\emph{729-743}, Feb 2017. .

\bibitem{b2} X. Liu, Z. Xu, L. Wang, W. Dong and S. Xiao, “Cognitive Dwell Time Allocation for Distributed Radar Sensor Networks Tracking via Cone Programming,” \emph{IEEE Sensors Journal}, vol. $20$, no. $10$, pp. \emph{5092-5101}, May 2020.

\bibitem{b3} J. Yan, W. Pu, H. Liu, S. Zhou, and Z. Bao, “Cooperative target assignment and dwell allocation for multiple target tracking in phased array radar network,” \emph{Signal Processing}, vol. $141$, pp.\emph{74-83}, 2017.

\bibitem{b5} Y. Zhang, and G. Shan, “A Risk-Based Multisensor Optimization Scheduling Method for Target Threat Assessment,” \emph{Mathematical Problems in Engineering}, vol. $2019$, 2019.

\bibitem{b6} F. Bolderheij, F. G. J. Absil, and P. Genderen, “A risk-based object-oriented approach to sensor management.”, in \emph{2005 7th International Conference on Information Fusion.}, , 2005.

\bibitem{b11} W. Jundi, X. Yunshan, X. bingsong, X. haibao, “Sensor-target assignment algorithm based on complementary principle.”, in \emph{2018 Chinese Automation Congress (CAC)}, 2018, pp 1774-1777.

\bibitem{b13} M. Zhiying,T. Shujuan, D. Qinglin, W. Songqiao, W. Nan, “Airborne sensor management algorithm based on targets threat evaluation.” in \emph{2017 13th IEEE International Conference on Electronic Measurement \& Instruments (ICEMI).}, 2017.

\bibitem{b14}A. Bab-Hadiashar, and W. Liu, “Sensor-management for multitarget filters via minimization of posterior dispersion,” \emph{IEEE Transactions on Aerospace and Electronic Systems}, vol. $53$, no. $6$, pp.\emph{2877-2884}, Dec 2017..

\bibitem{b23} S. Frost, K. Goebel, and J. Celaya, “A Briefing on Metrics and Risks for Autonomous Decision-making in Aerospace Applications,” in \emph{AIAA Infotech at Aerospace Conference and Exhibit 2012}, 2012, pp.2402.

\bibitem{b7} H. Zhang, J. Xie, B. Zong, W. Lu, and C. Sheng, “Dynamic priority scheduling method for the air-defence phased array radar,” \emph{IET Radar, Sonar \& Navigation}, vol. $11$, no. $7$, pp.\emph{1140-1146}, 2017..

\bibitem{b12} B. Li, L. Tian, D. Chen, and Y. Han, “A task scheduling algorithm for phased-array radar based on dynamic three-way decision,” \emph{Sensors}, vol. $20$, no. $1$, pp.\emph{153}, 2020.

\bibitem{b8} C. Pang, and G. Shan, “Sensor scheduling based on risk for target tracking,” \emph{IEEE Sensors Journal}, vol. $19$, no. $18$, pp.\emph{8224-8232}, 2019.

\bibitem{b10} C. Pang, and G. Shan, “Risk-based sensor scheduling for target detection,” \emph{ Computers \& Electrical Engineering}, vol. $77$, pp.\emph{179-190}, 2019.

\bibitem{b9} F. Katsilieris, H. Driessen, and A. Yarovoy, “Threat-based sensor management for target tracking,” \emph{IEEE Transactions on Aerospace and Electronic Systems}, vol. $51$, no. $4$, pp.\emph{2772-2785}, 2015.
   
\bibitem{b28} A. Charish.(2015). Attention in Cognitive Radar Using Effective Resources Management[PDF]. Available :https://www.sto.nato.int/publications/STO

\bibitem{b30} S. Wang, Z. Liu, R. Xie, and L. Ran "Online Sequential Extreme Learning Machine-Based Active Interference Activity Prediction for Cognitive Radar.", \emph{Remote Sensing}, vol. $14$, no. $12$, 2022.

\bibitem{b31} K. Pattanayak, V. Krishnamurthy, and C. Berry, "Meta-cognition. an inverse-inverse reinforcement learning approach for cognitive radars." in \emph{2022 25th International Conference on Information Fusion (FUSION)}, 2022.

\bibitem{b32} T. D. Ridder, A. F. Martone, B. H. Kirk, and R. M. Narayanan, "Multiple criteria operational reliability performance metric of a metacognitive tracking radar"., \emph{IEEE Transactions on Aerospace and Electronic Systems}, 2023.

\bibitem{b29} Y. Shen, H. Wymeersch and M. Z. Win, “Fundamental Limits of Wideband Localization— Part II: Cooperative Networks,” \emph{IEEE Transactions on Information Theory}, vol. 56, no. 10, pp. \emph{4981-5000}, Oct. 2010. 

\bibitem{b19} M. Fatemi and S. Haykin, “Cognitive control: Theory and application,” \emph{IEEE Access}, vol. $2$, pp.\emph{698-710}, 2014.

\bibitem{b21} X. Liu, Z. Xu, W. Dong, L. Wang, and X. Li, “Cognitive Resource Allocation for Target Tracking in Location-Aware Radar Networks,” \emph{IEEE Signal Processing Letters}, vol. $27$, pp.\emph{650-654}, 2020.
\end{thebibliography}

\end{document}